\documentclass[prb,twocolumn,superscriptaddress,showpacs,unsortedaddress]{revtex4}
\usepackage{graphicx}
\usepackage{dcolumn}
\usepackage{amsmath}
\usepackage{hyperref}
\usepackage{bm}
\usepackage{color}
\usepackage{SIunits}

\newcommand{\kcro}{$\mathrm{K_2CrO_4 }$}

\begin{document}

\title{Disentangling multipole resonances through a full x-ray polarization analysis.}

\author{C. Mazzoli} 

\affiliation{European Synchrotron Radiation Facility, BP 220, F-38043
  Grenoble Cedex 9, France}

\author{S.B. Wilkins} 

\affiliation{European Synchrotron Radiation Facility, BP 220, F-38043
  Grenoble Cedex 9, France}
\affiliation{Brookhaven National Laboratory, Condensed Matter Physics
  \& Materials Science Department, Upton, NY, 11973-5000, USA}

\author{S. Di Matteo} 

\affiliation{Laboratori Nazionali di Frascati INFN, via E. Fermi 40,
  C.P. 13, I-00044 Frascati (Roma) Italy}
\affiliation{Equipe de physique des surfaces et interfaces, UMR-CNRS 6627 PALMS,
Universit\'e de Rennes 1, 35042 Rennes Cedex, France}

\author{B. Detlefs}

\affiliation{European Synchrotron Radiation Facility, BP 220, F-38043
  Grenoble Cedex 9, France}

\affiliation{European Commission, JRC, Institute for Transuranium
   Elements, Postfach 2340, Karlsruhe, D-76125 Germany}

\author{C. Detlefs}

\affiliation{European Synchrotron Radiation Facility, BP 220, F-38043
  Grenoble Cedex 9, France}

\author{V. Scagnoli}

\affiliation{European Synchrotron Radiation Facility, BP 220, F-38043
  Grenoble Cedex 9, France}

\author{L. Paolasini}

\affiliation{European Synchrotron Radiation Facility, BP 220, F-38043
  Grenoble Cedex 9, France}

\author{P. Ghigna}

\affiliation{Dipartimento di Chimica Fisica ``M. Rolla'', Universit\'a
  di Pavia, I-27100 Pavia, Italy}

\begin{abstract}
Complete polarization analysis applied to resonant x-ray scattering at the
$\mathrm{Cr}$ K-edge in \kcro\ shows that incident linearly polarized x-rays
can be converted into circularly polarized x-rays by diffraction at
the Cr pre-edge ($E=5994\,\mathrm{eV}$).
The physical mechanism behind this phenomenon is a subtle interference effect
between purely dipole (E1-E1) and purely quadrupole (E2-E2) transitions,
leading to a phase shift between the respective scattering amplitudes.
This effect may be exploited to disentangle two close-lying resonances
that appear as a single peak in a conventional energy scan, in this way allowing
to single out and identify the different multipole order parameters involved.
\end{abstract}


\date{\today}
\maketitle

\section{Introduction}

In the last 10 years resonant x-ray scattering (RXS) has
developed into powerful technique to obtain direct information about
charge, magnetic, and orbital degrees of freedom
\cite{Blume94,Murakami98,Murakami98b,Joly03,Wilkins03,Wilkins03b}. It combines
the high sensitivity of x-ray diffraction to long-range ordered
structures with that of x-ray absorption spectroscopy to local electronic
configurations.  In particular, the development of third generation
synchrotron radiation sources has made possible the detection of small
effects in electronic distribution, due to magneto-electric anisotropy
\cite{Kubota04} or to local chirality \cite{Goulon98,DiMatteo05}, that
can be related to the interference between dipole (E1) and quadrupole
(E2) resonances.  These pioneering studies paved the way to a new
interpretation of RXS experiments in terms of electromagnetic
multipoles of higher order than dipole, and led to the detection of
phase transitions characterized by order parameters (OP) of exotic symmetry
\cite{Paixao02,DiMatteo03,Arima05}.
Several theories have been developed, based on these higher-order OP,
to explain "anomalous" phase transitions. For example, in NpO$_2$ a proper interpretation
of the magnetic ground-state requires a primary OP at least octupolar order
\cite{Caciuffo03}, whereas in high-temperature cuprate superconductors,
the pseudogap phase has been interpreted in terms of parity and
time-reversal odd toroidal multipoles\cite{Varma97,Varma99}.
In several cases, though, the assignment of the multipolar origin to
a RXS signal is not clear \cite{Mannix99,Paixao02,Paolasini99}. The
characteristic variation of the intensity and polarization as the
sample is rotated about the scattering vector during an azimuthal scan
may allow a clearer identification of the order of the
multipole. However, this technique is plagued by many experimental
difficulties either from the sample, e.g. when the crystal presents
twinning and mosaicity, or due to restrictive sample environments, such as
cryomagnets. Moreover, it is very difficult to identify and resolve
two resonances of different multipolar origin when they are separated
by less than $\sim 1\,\mathrm{eV}$, a situation which frequently
occurs at the metal pre K-edge region of transition metal oxides, where E1 and E2
transitions can have similar magnitudes, or at L$_{2,3}$
edges of rare-earth compounds. 
For example, at the Fe pre K-edge in $\alpha$-Fe$_2$O$_3$,
evidence of a E2-E2 transition was found already in 1993
\cite{Finkelstein92}. This transition was interpreted as the signature of an
hexadecapolar electronic ordering\cite{Carra94}. Later, however, it was shown by
symmetry arguments that also an axial-toroidal-quadrupole OP, parity-breaking,
was hidden within the same resonance \cite{DiMatteo03}, but to date no conclusive
experimental evidence for this interpretation has been provided. 
Analogously, at the Ce L$_{2,3}$-edges in CeFe$_2$ the different electronic
OP related to $4f$ and $5d$ states are entangled and cannot be examined
individually by conventional RXS \cite{Paolasini07b}.
A further example is provided by \kcro. Its space group symmetry ({\it Pnma}, No. 62) 
allows several excitation channels at the (1k0) Bragg forbidden reflections \cite{Templeton94},
due to the presence of multiple moments of different symmetries
(electric quadrupole, octupole, hexadecapole) in the same energy range, as described
in more details below.

The aim of the present article is to address the previous problems by extending the well-known techniques
of optical polarimetry from the visible to the x-ray regime, as developed
at the beamline ID20\cite{ID20} at the ESRF, Grenoble, France.
By using a diamond x-ray phase plate to rotate the incident linear
polarisation in combination with a linear polarisation analyser,
we can resolve resonances determined by multipoles of different
order that are very close in energy, playing on their relative phase shifts. 
The idea can be explained through a simplified model, by considering two externally driven,
damped harmonic oscillators of unitary amplitude, but with resonant
frequencies differing by $2\zeta$.
The scattering amplitude of such oscillators is given by
\begin{equation}
g_{\pm} (\omega) = \frac{1}{\omega \pm \zeta - i\Gamma}.
\label{toymodel}
\end{equation}
Here $\Gamma$ is the inverse damping time, and $\omega$ the frequency of
the external excitation. We also suppose that the two resonances scatter
in different polarization channels. Using the Jones matrix formalism
\cite{Blume88} the polarisation of the scattered beam may then be written
as $\epsilon'=G \epsilon$, where $\epsilon$ ($\epsilon'$) are the
incident (scattered) polarisation vectors, and the matrix 
$G=\left(\begin{array}{cc}
g_- & g_+ \\ g_+ & g_-
\end{array}\right)$
contains the dependence on the photon energy.
Experimentally, the scattered beam polarisation is best described in terms of
the Poincar\'e-Stokes parameters:

\begin{eqnarray}
    P'_1  & \equiv &   \frac{|\epsilon_{\sigma}'|^2-|\epsilon_{\pi}'|^2}{P'_0} \nonumber  \\
    P'_2  & \equiv &    2\Re e\frac{\epsilon_{\sigma}'^{*}\epsilon_{\pi}'}{P'_0} \nonumber   \\
    P'_3  & \equiv &   2\Im m\frac{\epsilon_{\sigma}'^{*}\epsilon_{\pi}'}{P'_0} \nonumber
   \label{stokesth} 
\end{eqnarray}

with $P'_0\equiv (|\epsilon_{\sigma}'|^2+|\epsilon_{\pi}'|^2)$ the total intensity,
and $\epsilon'^{*}$ the complex conjugate of $\epsilon'$.
$P'_1$ and $P'_2$ describe the linear polarization states, whereas $P'_3$ indicates the degree
of the circular polarization. The Poincar\'e-Stokes parameters, $P_{0,1,2,3}$ of the incident
beam are obtained by substituting $\epsilon$ for $\epsilon'$.

For example, for an incoming $\pi$ polarised beam, $\epsilon=\left(\begin{array}{c}0\\1\end{array}\right)$,
we obtain $\left(\begin{array}{c}
\epsilon_{\sigma}'\\\epsilon_{\pi}'\end{array}\right)=\left(\begin{array}{c}g_+\\g_-\end{array}\right)$,
which in turn yields:


\begin{eqnarray}
    P'_1  & = &   - \frac{2 \zeta \omega}{\omega^2+\zeta^2+\Gamma^2} \nonumber  \\
    P'_2  & = &   + \frac{\omega^2-\zeta^2+\Gamma^2}{\omega^2+\zeta^2+\Gamma^2}    \\
    P'_3  & = &   + \frac{2\zeta\Gamma}{\omega^2+\zeta^2+\Gamma^2} \nonumber
   \label{stokestoy} 
\end{eqnarray}

\begin{figure}
  \includegraphics[width=\columnwidth]{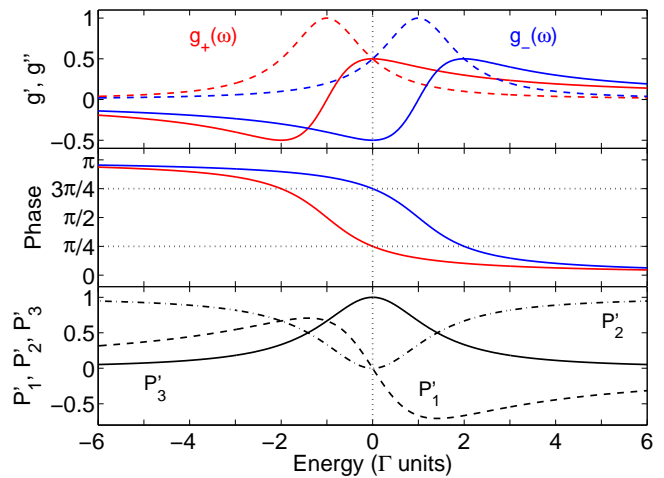}
  \caption{(Color) - Top: Typical behavior of real ($g'$, continuous) and imaginary
($g''$, dashed) parts for the two resonators of Eq.~\ref{toymodel}
($g_{+}$ red; $g_{-}$ blue), with $\zeta=\Gamma$.
Middle: Corresponding phase shifts. Bottom: Poincar\'e-Stokes parameters
as calculated from Eqs.~\ref{stokestoy}.} 
  \label{fig_toymodel}
\end{figure}

\noindent
Therefore we expect that in the intermediate region between the
two resonances, the outgoing beam is circularly polarized, depending on the relative
dephasing of the two resonances, as shown in
Fig.~\ref{fig_toymodel} for the case $\zeta=\Gamma$. 

Below, we describe experimental data which we then compare to quantitative {\it ab-initio}
calculations carried out using the FDMNES code \cite{Joly01}.
We demonstrate that a $100\%$ linear- to circular-polarization conversion at the
pre-edge region of the Cr K-edge in \kcro\ is induced by the interference of
the dispersive and absorptive parts of two different multipoles probed by purely dipole (E1-E1)
and purely quadrupole (E2-E2) resonances. 
Thus, the scattered beam originates from two different excitation channels, each scattering
the beam with a different phase.

Their relative amplitudes, at a given energy, are governed by the probed multipoles,
the relative orientation of the incident and scattered electric field vectors to
the probed multipoles and the reciprocal lattice point under study.
Their interference can therefore be tuned simply by varying the incident polarization by
means of a phase plate. 

\section{Experimental setup}

\begin{figure}
  \includegraphics[width=0.9\columnwidth]{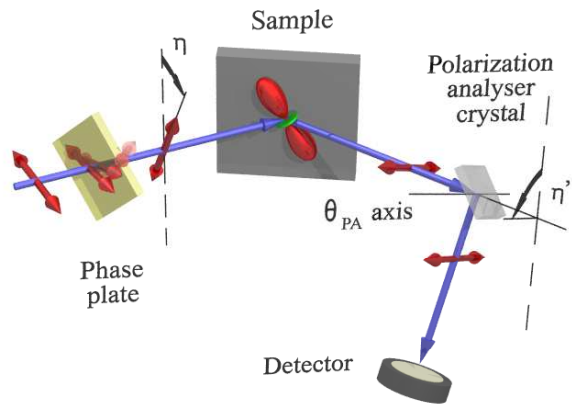}
  \caption{(Color online) Experimental setup with phase plate in half-wave mode;
x-ray directions indicated by blue arrows, polarizations by red ones.
Synchrotron light arrives from the left, horizontally polarized ($\pi$).
$\eta$ is the rotation angle of the incident polarization.
$\eta'$ is the rotation angle of the polarization analyser crystal;
the zero positions of the two angles, corresponding to $\sigma$ and $\sigma'$
polarisations respectively, are represented by dashed lines.
The continuous line is the the rocking axis of the polarisation analysis crystal
$\theta_{\mathrm{PA}}$. The polarisation analyser stage is shown in the configuration corresponding
to the maximum detected intensity.} 
  \label{fig_exp_setup}
\end{figure}

Experiments were carried out at ID20, ESRF. The experimental setup is outlined
in Fig.~\ref{fig_exp_setup}.  A single crystal of \kcro\ was
mounted on the six-circle horizontal diffractometer, and a cryostat was used
to stabilize the temperature at 300~K.
Sample rocking curves ($\theta$-scans) of the
crystal resulted in widths smaller than $0.01^{\circ}$ indicating a
high sample quality.

A diamond phase plate of thickness 700~\micro\meter\ with a [110]
surface was inserted into the incident beam, within its own
goniometer, and the (111) Bragg reflection in symmetric Laue geometry 
was selected to modify the
polarization of the x-ray beam incident on the diffractometer. The
phase plate was operated in either quarter-wave or half-wave plate mode.
With the former we generated left- or right-circular
polarization, $P_3 \approx \pm 1$, whereas with the latter we rotated the linear
incident polarization into an arbitrary plane \cite{Giles95,Bouchenoire06}, 
described by $P_1 \approx \cos(2\eta)$ and $P_2 \approx \sin(2\eta)$. Here $\eta$ is 
the angle between the incident beam electric field vector and the vertical axis 
(see Fig.~\ref{fig_exp_setup}), i.e. $\eta=0$ when the polarization 
is perpendicular to the horizontal scattering plane ($\sigma$ polarization).

\begin{figure}
  \includegraphics[width=0.9\columnwidth]{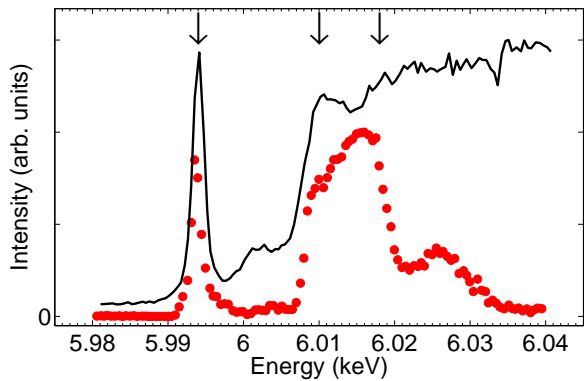}
  \caption{(Color online) Experimental data on fluorescence yield (black continuous line)
and energy scan (red symbols) for the (130) reflection in $\pi\rightarrow\sigma'$ polarization
configuration. Arrows indicate the energy values where Stokes' parameters were measured.} 
  \label{fig_spectrum}
\end{figure}

\begin{figure}
  \includegraphics[width=0.9\columnwidth]{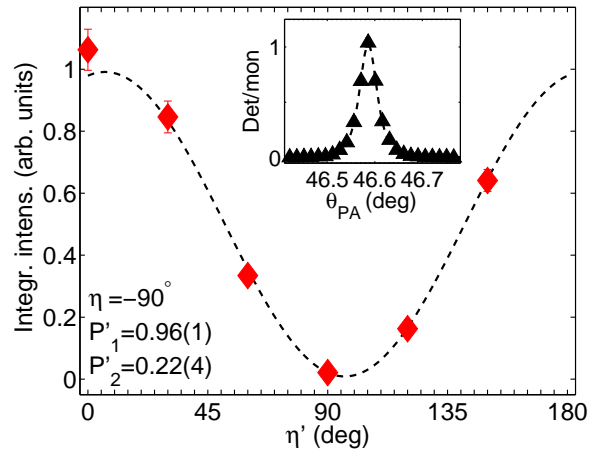}
  \caption{(Color online) Variation of the integrated intensity of the (130) forbidden
Bragg peak as a function of the polarization analyzer angle $\eta'$. The dashed line
represents a fit to these data with Eq.~\ref{stokespar} (see text for details).
The insert is an example of a rocking scan of the analyser: the integrated instensity
obtained from the fit (dashed line: lorentian squared and linear background)
represents experimental point in the main axes.} 
  \label{fig_single_stokes}
\end{figure}

The sample was mounted with the [010] and [100] directions defining
the horizontal scattering plane.
Figure~\ref{fig_spectrum} shows the fluorescence yield and the energy dependence
of the glide-plane forbidden (130) reflection, collected at an azimuthal angle of
$\Psi=-0.78(3)^{\circ}$ degrees with respect to the reciprocal lattice direction [010].
The photon energy was then tuned to the pre-edge region of the Cr K-edge
($5994\,\mathrm{eV}$).
An x-ray polarization analyzer was placed in the scattered beam. It exploited the 
(220) Bragg reflection of a LiF crystal, scattering close to Brewster's angle of $45^{\circ}$.
The polarization analyzer setup was rotated around the scattered beam by an angle,
$\eta'$, and at each point the integrated intensity was determined by rocking
the analyzer's theta axis ($\theta_{\mathrm{PA}}$). An example is shown in the inset of
Fig.~\ref{fig_single_stokes}. The resulting integrated intensities
were then fitted to the equation
\begin{equation}
  I = \frac{P'_{0}}{2}\left[ 
    1 + P'_{1}\cos 2\eta' + P'_{2}\sin 2\eta' 
    \right],
\label{stokespar} 
\end{equation}
to obtain the Poincar\'e-Stokes parameters, $P'_1$ and $P'_2$.
An example is shown in Fig.~\ref{fig_single_stokes} for $\eta = $~-90$^\circ$. 
The degree of circular polarization, $P'_{3}$, can not be
measured directly in this setup. However, an upper limit is inferred from
$P'^2_1+P'^2_2+P'^2_3 \leq 1$ (the equality holding for a completely polarized beam).

\begin{figure}
  \includegraphics[width=0.9\columnwidth]{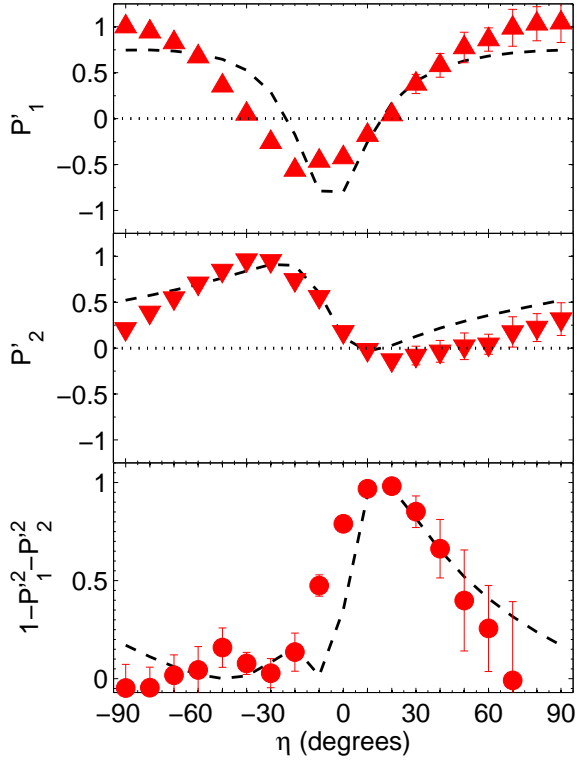}
  \caption{(Color online) Calculated (dashed lines)
and measured (symbols) Stokes' parameters for $(130)$ reflection versus
the polarisation angle of the linear incoming light (see Fig.~\ref{fig_exp_setup}),
$E=5.994\,\mathrm{keV}$. See text for details.}
  \label{fig_polscan}
\end{figure}

\begin{figure}
  \includegraphics[width=0.9\columnwidth,angle=0]{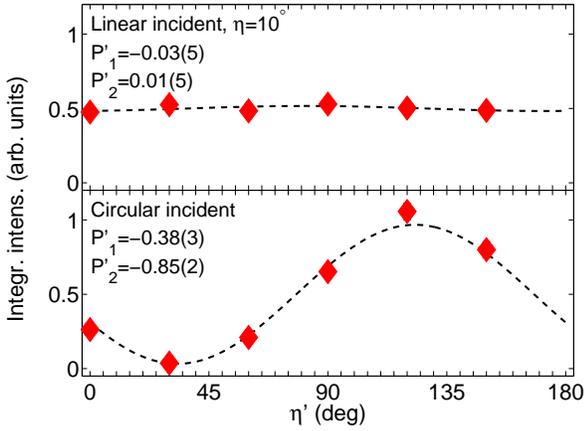}
  \caption{(Color online) Stokes' parameters for $(130)$ reflection with (top) linearly
($\eta$=10$^{\circ}$) and (bottom) circularly polarized incident x-rays, collected at
$E=5.994\,\mathrm{keV}$. Dashed lines are fit with Eq.~\ref{stokespar}.}
  \label{fig_reverse_path}
\end{figure}

We systematically measured $P'_1$ and $P'_2$ of the beam scattered at
the (130) reciprocal lattice point as function of $\eta$.
Figure~\ref{fig_polscan} shows both the experimental data (symbols) and
the theoretical calculation (dashed lines), described below.
The measured degree of linear polarization of the scattered beam, $P'^2_1+P'^2_2$,
strongly deviates from 100\% in the range $-10^\circ \lesssim \eta \lesssim 50^\circ$,
indicating that a large component of the scattered beam is either
circularly polarized or depolarized.
To ascertain which of these two processes is realised we reconfigured the diamond phase plate
to produce circularly polarized x-rays. Figure~\ref{fig_reverse_path} shows  the measured
$P'_1$ and $P'_2$ for both linearly and circularly polarized x-rays. The presence of linearly
scattered x-rays for the circular incident case is consistent with the assumption that
the region for which $P'^2_1+P'^2_2$ strongly deviates from 1 corresponds to an increased
circularly polarized contribution.
Furthermore, the calculations for $P'^2_3$, detailed in Section III,
involving only completely polarized contributions to the scattered beam,
agree well with its upper limit inferred from the data ($P'^2_3 \equiv 1-P'^2_1-P'^2_2$),
indicating that the signal is essentially circularly polarised.

\section{Theoretical discussion}

In RXS the global process of photon absorption, virtual photoelectron
excitation and photon re-emission, is coherent throughout the crystal,
thus giving rise to the usual Bragg diffraction condition
$\sum_{j}e^{i\vec{Q}\cdot\vec{R}_j}(f_{0j}+f_j'+if_j'')$.
Here $\vec{R}_j$ stands for the position of the scattering center $j$,
$\vec{Q}$ is the diffraction vector and $f_{0j}$ is the Thomson factor.
$f_j'$ and $f_j''$, related by
Kramers-Kronig transform, are given, at resonance, by the
expression \cite{Blume94}:

\begin{equation}
  f_j'+if_j''
  \equiv 
  f_j(\omega) 
  \propto -\omega^2
  \sum_{n} \frac{
    \langle \psi_g(j)|\hat{O}'^* |\psi_n \rangle
    \langle \psi_n|\hat{O} |\psi_g(j) \rangle
  }{
    \omega-(\omega_n-\omega_g)-i\frac{\Gamma_n}{2}
  }, 
  \label{eq_arxs}
\end{equation}
where $\omega$ is the photon energy, $\omega_g$ the ground state
energy, $\omega_n$ and $\Gamma_n$ are the energy and inverse lifetime of
the excited states, $\psi_g(j)$ is the core ground state centered
around the $j^{th}$ atom and $\psi_n$ the photo-excited state, $\epsilon$ and $\epsilon$' are the
polarizations of the incoming and outgoing photons and $\vec{q}$ and $\vec{q'}$ their
corresponding wave vectors.
The sum is extended over all the excited states of the system. The transition
operator $\hat{O}^{(')}= \vec{\epsilon}^{(')} \cdot \vec{r} ~ \big(1
- \frac{i}{2}\vec{q}^{(')} \cdot \vec{r}\big)$ is written as a
multipolar expansion of the photon field up to electric dipole (E1)
and quadrupole (E2) terms; $\vec{r}$ is the electron position relative to
the resonating ion, $\vec{\epsilon}^{(')}$ is the polarization of
the incoming (outgoing) photon and $\vec{q}^{(')}$ its corresponding
wave vector.

By taking into account the space group symmetry of \kcro\
({\it Pnma}, No. 62), the four equivalent Cr sites at Wyckoff $4c$ positions
(with local mirror-plane ${\hat{m}}_y$) can be related one another by
the following symmetry operations: $f_3={\hat{I}}f_1$,
$f_4={\hat{C}}_{2x}f_1$ and $f_2={\hat{I}}f_4$.  ${\hat{I}}$ is the
space-inversion operator and ${\hat{C}}_{2x}$ is the $\pi$-rotation
operator around $x$-axis.  The Thomson scatering $f_{0j}$ does not
contribute at $(1k0)$ type reflections, for any k, due to the glide plane extinction rule.
Therefore, the structure factor at Cr K-edge for the $(1k0)$ chosen reflections,
when summed over all equivalent sites, becomes, for E1-E2 scattering:
\begin{equation}
  S = 2i\sin[2\pi(x+\frac{k}{4})] (1- {\hat{m}}_{z}) f_1
  \label{strucfac1}
\end{equation}
while for E1-E1 and E2-E2 scattering it is given by:
\begin{equation}
  S = 2\cos[2\pi(x+\frac{k}{4})] (1- {\hat{m}}_{z}) f_1.
  \label{strucfac2}
\end{equation}
Here $x\simeq 0.23$ is the fractional coordinate of Cr atoms and
${\hat{m}}_z$ is a glide-plane orthogonal to the $z$-axis. In deriving
Eqs. \ref{strucfac1} and \ref{strucfac2}, we have used the
identity $f_1={\hat{m}}_y f_1$. It is interesting to note the
different behavior of the two terms for $k$ even or odd. For example, when
$k=4$, the E1-E2 scattering is proportional to $\sin(2\pi x)\simeq
0.99$ and it dominates the other terms, proportional to $\cos(2\pi
x)\simeq 0.12$.  Indeed we found the presence of a very intense pre-edge
feature from E1-E2 channel at the (140) reflection  that is related to
the electric octupole moment, as predicted in Ref. \onlinecite{Marri04} and
verified numerically by our {\it ab-initio} calculations.
However, as described above, the presence of a single
scattering channel, as in the case of the (140) reflection, can not lead to a
circularly polarized diffracted beam.
This can be demonstrated by a symmetry argument: if only one scatterer is present,
which is by hypotesis non-magnetic, and the incident light is linearly polarized,
then the initial state is time-reversal even. Therefore, as matter-radiation interaction
does not break time-reversal, it follows that the final state must also be time-reversal even,
i.e., radiation can not be circularly polarized, which would break time-reversal.
This is no more true when two scatterers are present, due to the extra degree of
freedom represented by the time (phase) delay between the two scattering processes. 
Indeed, this was {\it a posteriori} verified by our numerical simulation,
which confirmed that no outgoing circular polarization is present at the (140) reflection.

The case of the (130) reflection is very different. The role of
$\sin(2\pi x)$ and $\cos(2\pi x)$ in Eqs. \ref{strucfac1} and
\ref{strucfac2} switches in such a way that the two
diffraction channels E1-E1 and E2-E2 become predominant. Further analysis
of the structure factor\cite{note1} reveals that two resonances are allowed,
one for each channel, corresponding to an
electric quadrupole ordering for the E1-E1 scattering and an electric
hexadecapole \cite{Carra94} for the E2-E2 scattering.
Finally, multiple-scattering calculations with the FDMNES program
confirm that the two resonances overlap in the pre-edge region, though
slightly shifted in energy of $\sim 1$ eV.
These are the conditions to be met to get the interference of the two channels. 
In order to describe the effect quantitatively from a
theoretical point of view we used the {\it ab-initio}
code FDMNES, in the multiple-scattering mode, to
calculate $P_3$ directly from Eq. \ref{eq_arxs} for the \kcro\
structure\cite{Templeton94}. We employed a cluster of 43 atoms, corresponding to
a radius of 5.5~\AA~around the resonating Cr-atom. Notice that in this most general case we find
$P'_3(\omega)\propto\left(f'_{E1}(\omega)f''_{E2}(\omega)-f'_{E2}(\omega)f''_{E1}(\omega) \right)$
where $f'$ and $f''$ are the usual dispersive and absorptive
terms (see Eq. \ref{eq_arxs}) for E1 and E2 channels. Therefore at the photon
energy $\omega$, $P'_3$ is determined by the interference of the absorptive part
$f''$ of one channel with the
dispersive part $f'$ of the other and, again, in the presence of just
one channel, $P'_3=0$.

The numerical simulations shown in
Fig.~\ref{fig_polscan} (dashed lines) confirm that
there is an azimuthal region where the incoming linear polarization is fully
converted into an outgoing circular polarization and that
the effect is determined by the interference of the E1-E1 and E2-E2
channels.

In this respect, it is interesting to note that the origin
of this effect is profoundly different from those determined by a
chiral (magnetic) structure (see, e.g.,
Ref. \onlinecite{Sutter97,  Lang04}), as clearly seen by the fact that
all the tensors involved are non-magnetic and parity-even.

Finally, we verified experimentally that at the Cr K-edge ($E=6010$ and $6018\,\mathrm{eV}$),
where only one term in the E1-E1 channel is present,
no circular polarization was observed for all incident angles, i.e., $P'^2_1+P'^2_2=1$.
Calculations performed using FDMNES confirmed this result.

\section{Conclusions}

Polarization analysis of RXS experiments has developed greatly in the
last few years, helping to understand several characteristics of
order parameters in transition metal oxides, rare-earth based compounds, and actinides.
Up to now, however, the full investigation of Stokes' parameters was
not applied most likely because linear polarization analysis, where only the $P'_1$
parameter was determined by measuring the $\sigma\rightarrow\sigma'$ and
$\sigma\rightarrow\pi'$ channels, was considered sufficient.
While this may be true when just one excitation channel is involved (as at the
(140) reflection in the present case), several dephasing phenomena may
appear when two different multipole excitations close in energy are
involved in the transition. As we have seen, these phenomena may lead
to a situation where incoming linear polarization is scattered to a
circular polarization due to an interference between two multipoles, at the same
time allowing for a very sensitive determination of the presence of the
second  multipole.
We believe that the use of phase plates and of a complete polarization analysis,
is the key to disentangle multi-resonance structures in those situations where an
usual energy scan, like the one shown in Fig.~\ref{fig_spectrum},
 is not sufficient to this aim. Effect of {\it d}-band filling on details of
the electronic structure will be investigated by the method presented here
in the series of isostructural compounds
\kcro\ $\rightarrow \mathrm{K_2MnO_4} \rightarrow \mathrm{K_2FeO_4}$.

\begin{acknowledgments}

The authors would like to acknowledge David H. Templeton
and Fran\c cois de Bergevin for enlightening discussions.

One of us (SDM) acknowledges the kind hospitality at ESRF during the
preparation of the manuscript.

The work at Brookhaven National Laboratory is supported by the U.S.  
Department of Energy, under contract no. DE-AC02-98CH10886.

\end{acknowledgments}

\bibliography{mazzoli}

\begin{thebibliography}{30}
\bibitem[{not()}]{note1}
\emph{\bibinfo{title}{The application of the structure-factor symmetry
  operation ${\hat{m}}_z$ on the irreducible tensor representing the atomic
  scattering factor \cite{DiMatteo03} leads to ${\hat{m}}_z
  f^{(r)}_q=(-)^{(q+P)}f^{(r)}_q$, where $p$ is the parity of the tensor (+1
  for e1-e1 and e2-e2, -1 for e1-e2). from this rule the allowed multipole
  components follow.}}
\end{thebibliography}

\end{document}